\documentclass[pra, aps, twocolumn, floatfix, showpacs]{revtex4}
\usepackage{graphicx, amsmath, amssymb, times}

\begin{document}
\title{Quantum phases of atomic Fermi gases with anisotropic spin-orbit coupling}
\author{M. Iskin$^1$ and A. L. Suba{\c s}{\i}$^2$}
\affiliation{
$^1$Department of Physics, Ko\c c University, Rumelifeneri Yolu, 34450 Sar{\i}yer, Istanbul, Turkey. \\
$^2$Department of Physics, Faculty of Science and Letters, Istanbul Technical University, 34469 Maslak, Istanbul, Turkey. 
}
\date{\today}

\begin{abstract}

We consider a general anisotropic spin-orbit coupling (SOC) and analyze the phase 
diagrams of both balanced and imbalanced Fermi gases for the entire 
BCS--Bose-Einstein condensate (BEC) evolution. In the first part, we
use the self-consistent mean-field theory at zero temperature, and show that 
the topological structure of the ground-state phase diagrams is quite robust 
against the effects of anisotropy. In the second part, we go beyond the 
mean-field description, and investigate the effects of Gaussian fluctuations 
near the critical temperature. This allows us to derive the time-dependent 
Ginzburg-Landau theory, from which we extract the effective mass of the Cooper 
pairs and their critical condensation temperature in the molecular BEC limit.

\end{abstract}

\maketitle

\section{Introduction}
\label{sec:intro}

The realization of BCS--Bose-Einstein condentate (BEC) evolution 
with two-component atomic Fermi gases has received tremendous 
attention in the last decade~\cite{review, rmp}. In these experiments,
the tuning of attractive interactions permits the ground state 
of the system to evolve from a weak fermion attraction BCS limit 
of loosely bound and largely overlapping Cooper pairs to a 
strong fermion attraction limit of tightly bound small bosonic
molecules which undergo BEC.
The main difference between the BCS-BEC evolution problem
and the simple BCS theory is that the Cooper pairing is not 
allowed only for fermions with energies close to the Fermi energy 
but is also allowed for all momenta. 

All of the early theoretical works were concentrated on the balanced Fermi gases, 
i.e. both components have the same number and mass, suggesting 
that the evolution is not a phase transition but a smooth crossover,
and hence the name \textit{BCS-BEC crossover}. This prediction
was then found to be in very good agreement with all of the 
observations~\cite{review, rmp}. Motivated by the experimental success 
with balanced Fermi gases, many of the recent theoretical works 
were concentrated on imbalanced (population, mass and/or dimension) 
Fermi gases, suggesting various phases and phase transitions between them.
Some of these predictions were also confirmed by the following
experiments~\cite{imb1, imb2, imb3, imb4}, opening the door for 
new studies on more complicated systems with the hope of finding 
exotic phases of matter.

Arguably, one of the very promising new systems to investigate 
is the spin-orbit coupled (SOC) atomic Fermi gases~\cite{sato, kubasiak}. 
This is mainly motivated by the very recent success in realizing SOC atomic 
BEC~\cite{nist1, nist2}, and by a practical proposal for generating 
a SOC Fermi gas with $^{40}$K atoms~\cite{sau}. Some of the very recent
results on these systems can be summarized as follows. 
For the two-body problem, it has been found that a two-body 
bound state exists for some types of SOC, e.g. Rashba or Dresselhaus types, 
even on the BCS side ($a_s < 0$) of a resonance~\cite{shenoy1} 
with arbitrarily small $a_s \to 0^-$, 
where $a_s$ is the $s$-wave scattering length. For the many-body problem, 
it has been found for balanced Fermi gases that the SOC increases 
the single-particle density of states, which in return favors the 
Cooper pairing so significantly that increasing the SOC, while $a_s$ 
is held fixed, eventually induces a BCS-BEC evolution even for a 
weakly-interacting system when $a_s \to 0^-$~\cite{shenoy2, zhai, hui, han, chen}.
Similar to what happens in the usual BCS-BEC crossover picture 
of a balanced Fermi gas with increasing $1/a_s$, the BCS-BEC evolution 
with increasing SOC turns out to be a smooth crossover but not a 
phase transition. Therefore, the ground state of balanced Fermi 
gases with or without the SOC is a topologically trivial gapped superfluid.

However, the BCS-BEC evolution with Rashba-type SOC is found to become 
a phase transition for population-~\cite{gong, iskin1, wyi} and/or 
mass-imbalanced~\cite{iskin2} Fermi gases. It has been found that the SOC 
counteracts the population imbalance, and that this competition tends 
to stabilize the uniform topologically nontrivial gapless superfluid phases 
against the phase separation. In addition, topological phase transitions 
associated with the appearance of momentum space regions with 
zero quasiparticle/quasihole energies have been found, the 
signatures of which could be observed in the momentum distribution 
or the single-particle spectral function~\cite{iskin1, wyi}. 

The anisotropic (in momentum space) nature of the SOC is also found 
to stabilize exotic superfluid phases. For instance, 
in sharp contrast to the no-SOC case where only the gapless superfluid 
phase supports population imbalance, both the gapless and gapped 
superfluid phases are found to support population imbalance in 
the presence of a Rashba-type SOC~\cite{iskin1, wyi}. Similarly, again in sharp 
contrast to the no-SOC case where only the gapped superfluid 
phase supports population balance, both the gapped and gapless 
superfluid phases are found to support population balance in 
mass-imbalanced SOC Fermi gases when the mass 
difference becomes large enough~\cite{iskin2}.

In this paper, we extend our recent works~\cite{iskin1, iskin2}, and 
study the effects of anisotropic SOC on the phase diagrams of both 
balanced and imbalanced Fermi gases throughout the entire BCS-BEC evolution. 
We analyze both zero and finite temperature phase diagrams, and 
the paper is organized as follows. First, we review the noninteracting and 
interacting two-body problem in Sec.~\ref{sec:tbp}, and calculate the binding 
energy of the two-body bound-state in vacuum. Second, we study the many-body 
problem in Sec.~\ref{sec:mbp}, where we derive the mean-field theory at zero temperature,
and use it to analyze the ground-state phase diagrams of imbalanced Fermi gases. 
Then, we investigate the Gaussian fluctuations in Sec.~\ref{sec:fluct} near 
the critical temperature, and calculate the effective mass of the Cooper pairs 
and their critical condensation temperature in the molecular BEC limit. 
Last, our conclusions are briefly summarized in Sec.~\ref{sec:conc}.

\section{Two-Body Problem}
\label{sec:tbp}

Before presenting our new results for the many-body problem, let us first introduce 
the model Hamiltonian and review some of the recent results for the noninteracting 
and interacting two-body problem.

For the noninteracting SOC fermions, the two-body Hamiltonian (in units of $\hbar = 1 = k_B$)  
can be written as
\begin{align}
\label{eqn:H0}
H_0 = \sum_{\mathbf{k}} \psi_\mathbf{k}^\dagger 
 \left( \begin{array}{cc}
\epsilon_{\mathbf{k},\uparrow} + S_{k_z} & S_\mathbf{k_\perp} \\
S_\mathbf{k_\perp}^* & \epsilon_{\mathbf{k},\downarrow} - S_{k_z}
\end{array} \right)
\psi_\mathbf{k},
\end{align}
where
$
\psi_{\mathbf{k}}^\dagger = [a_{\mathbf{k},\uparrow}^\dagger, a_{\mathbf{k},\downarrow}^\dagger]
$
with $a_{\mathbf{k},\sigma}^\dagger$ 
($a_{\mathbf{k},\sigma}$) creates (annihilates) a spin-$\sigma$ fermion with 
momentum $\mathbf{k} = (k_x, k_y, k_z)$, $\epsilon_{\mathbf{k},\sigma} = k^2/(2m)$ is the kinetic 
energy, and $S_\mathbf{k_\perp} = \alpha_x k_x - i \alpha_y k_y$ and $S_{k_z} = \alpha_z k_z$
are the spin-orbit fields with $\lbrace \alpha_x, \alpha_y, \alpha_z \rbrace \ge 0$. 
The eigenvalues of this Hamiltonian matrix are
\begin{align}
\label{eqn:ehel}
\varepsilon_{\mathbf{k},s} = \epsilon_{\mathbf{k},+} + s \sqrt{\left( \epsilon_{\mathbf{k},-}+S_{k_z} \right)^2 + |S_\mathbf{k_\perp}|^2},
\end{align}
where $s = \pm$ labels the helicity bands, and 
$
\epsilon_{\mathbf{k},s} = (\epsilon_{\mathbf{k},\uparrow} + s\epsilon_{\mathbf{k},\downarrow})/2
$ 
are the kinetic energy average and half of the kinetic energy difference of $\uparrow$ 
and $\downarrow$ fermions. The corresponding eigenfunctions 
$\mathbf{u}_s^\dagger = (u_{1,s}^*, u_{2,s}^*)$ are given by
$
u_{1,s}/u_{2,s} = S_\mathbf{k_\perp}/[\epsilon_{\mathbf{k},-} + S_{k_z} - s \sqrt{(\epsilon_{\mathbf{k},-} + S_{k_z})^2 + |S_\mathbf{k_\perp}|^2}].
$
Throughout this paper, we mainly consider four analytically tractable 
spin-orbit fields: 
(i) $\alpha_x  = \alpha$ and $\alpha_y = \alpha_z = 0$ corresponding to an equal 
mixture of Rashba-~\cite{rashba} and Dresselhaus-type~\cite{dresselhaus} SOC (ERD),
(ii) $\alpha_z = \alpha$ and $\alpha_x = \alpha_y = 0$ corresponding to a fully 
aligned SOC (FA),
(iii) $\alpha_x = \alpha_y = \alpha$ and $\alpha_z = 0$ corresponding to a purely 
Rashba- or Dresselhaus-type SOC (PRD), and
(iv) $\alpha_x = \alpha_y = \alpha_z = \alpha$ corresponding to a fully spherical 
SOC (FS). 
We note that ERD- and FA-type SOC are essentially the same for balanced Fermi
gases, but these cases differ substantially for imbalanced Fermi 
gases as we discuss below in Sec.~\ref{sec:mbp}.

For the interacting SOC fermions, it has recently been shown that the strength 
of the attractive particle-particle interaction $g \ge 0$ is related to the two-body 
binding energy $\epsilon_b \le 0$ in vacuum via,
$
1/g = (1/2) \sum_{\mathbf{k},s} 1/(2\varepsilon_{\mathbf{k},s} + \epsilon_{th} - \epsilon_b),
$
where $\epsilon_{th} = m\alpha^2$ is the energy threshold for the two-body bound state.
As usually done, the theoretical parameter $g$ can be eliminated in favor of the 
experimentally relevant $s$-wave scattering length $a_s$ via the relation,
$
1/g = -m V/(4\pi a_s) + \sum_\mathbf{k} 1/(2\epsilon_{\mathbf{k},+}),
$
where $V$ is the volume. 

\begin{figure} [htb]
\centerline{\scalebox{0.5}{\includegraphics{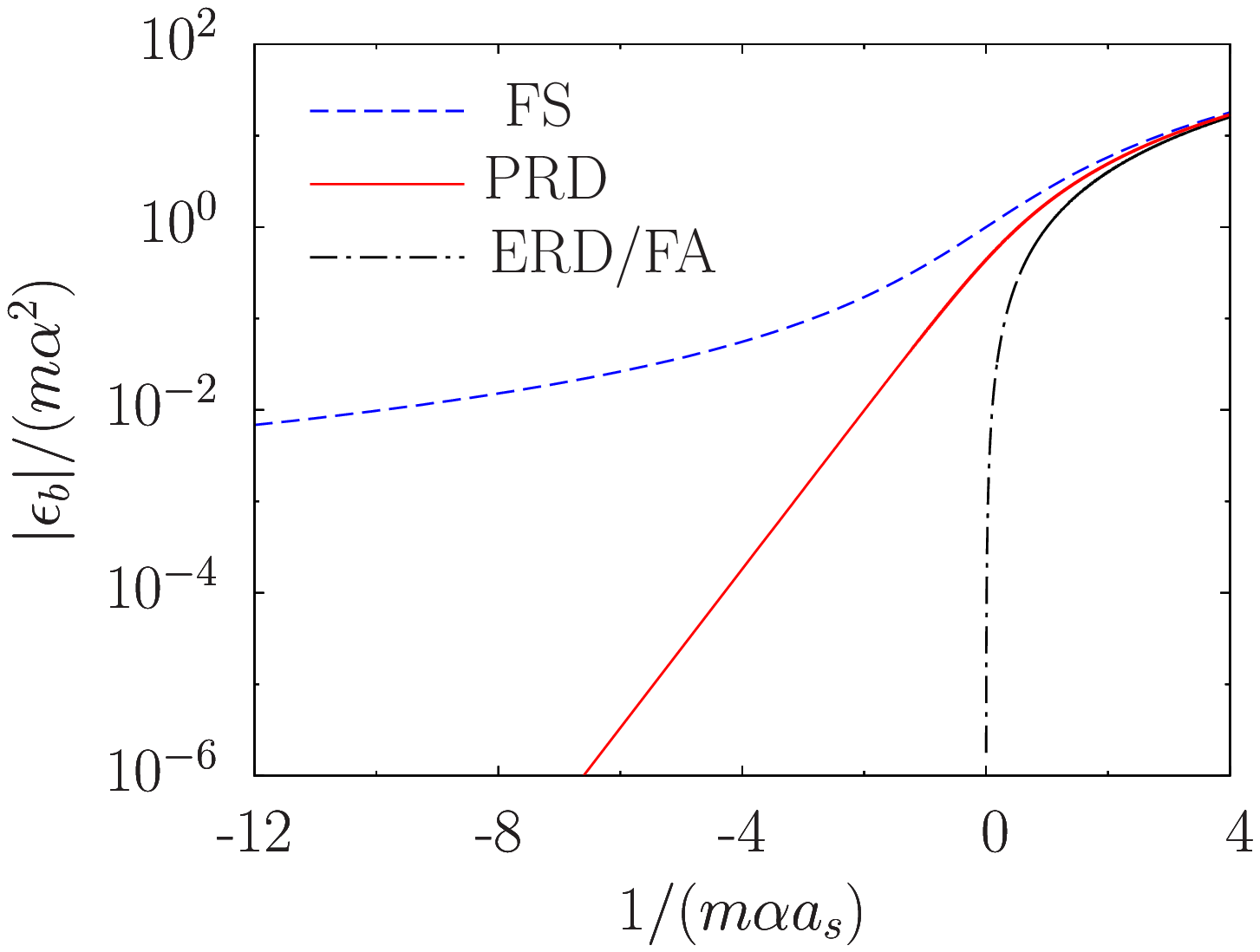}}}
\caption{\label{fig:eB} (Color online)
The binding energy $\epsilon_b \le 0$ of the two-body bound state are 
shown as a function $1/(m \alpha a_s)$ for the ERD-, FA-, PRD- and FS-type SOC.
}
\end{figure}

The bound-state equation is analytically tractable for the four 
cases mentioned above. For the simplest ERD- and FA-type SOC, a two-body bound 
state with energy $\epsilon_b = -1/(m a_s^2)$ exists only when $a_s > 0$, 
showing that the ERD- and FA-type SOC do not have any observable effect on 
the two-body problem.
On the other hand, in the case of PRD-type SOC, a two-body bound state exists even 
for $a_s < 0$~\cite{shenoy1}, and its energy is determined by~\cite{zhai, iskin1}
\begin{align}
\label{eqn:ebPRD}
\frac{1}{m \alpha a_s} = \sqrt{1 - \frac{\epsilon_b}{m\alpha^2}} 
- \ln \left( \sqrt{1-\frac{m\alpha^2}{\epsilon_b}}+\sqrt{\frac{m\alpha^2}{-\epsilon_b}} \right).
\end{align}
In the weak SOC limit, when $m \alpha^2 \ll |\epsilon_b|$, this expression gives
$
\epsilon_b \approx -1/(ma_s^2) -m\alpha^2 + 2\alpha/a_s
$
up to the leading order in $\alpha$, and it recovers the usual result in the $\alpha \to 0$ limit.
However, in the strong SOC limit, when $m \alpha^2 \gg |\epsilon_b|$, Eq.~(\ref{eqn:ebPRD}) 
gives
$
\epsilon_b \approx - (4m\alpha^2/e^2) e^{2/(m\alpha a_s)},
$
which is exponentially small as shown in Fig.~\ref{fig:eB}. 
We also obtain $\epsilon_b \approx -0.44 m\alpha^2$ at unitarity~\cite{zhai, hui}.
Similarly, in the FS-type SOC, an even deeper two-body bound state exists 
for $a_s < 0$, and its energy is given by
\begin{align}
\epsilon_b = -\frac{1}{2m a_s^2} - m\alpha^2 \pm \sqrt{\frac{1}{4m^2 a_s^4} + \frac{\alpha^2}{a_s^2}},
\end{align}
where $+$ ($-$) sign is valid for $a_s < 0$ ($a_s > 0$).
This expression reduces to 
$\epsilon_b \approx - m^3 \alpha^4 a_s^2$ in the weak- and
$\epsilon_b \approx -1/(m a_s^2) - 2m\alpha^2$ in the strong SOC limits, 
and to $\epsilon_b = -m\alpha^2$ at unitarity. 

Having shown that different types of SOC give rise to significant differences 
with regards to the existence of the two-body bound state and its binding energy, 
next we analyze the many-body problem where these differences also 
play an important role.

\section{Many-Body Problem}
\label{sec:mbp}

Let us first consider noninteracting ($g = 0$ or $a_s \to 0^-$) balanced 
($N_\uparrow = N_\downarrow = N/2$ or $\mu_\uparrow = \mu_\downarrow = \mu$) 
Fermi gases at zero temperature. For this purpose, 
and throughout this paper, we conveniently choose the energy (length) scale as the 
Fermi energy $\epsilon_F$ (Fermi momentum $k_F$) of $N/2$ fermions such 
that $N = k_F^3 V / (3\pi^2)$.

It has been shown that increasing the SOC for a noninteracting Fermi 
gas leads to a change in the Fermi surface topology, when the number of fermions 
in the $+$-helicity band ($N_+$) vanishes~\cite{shenoy2}, where 
$
N_s = \sum_\mathbf{k} \theta(\mu - \varepsilon_{\mathbf{k},s}).
$
This occurs when the chemical potential $\mu$ goes below the bottom of the energy 
band, i.e. when $\mu = 0$, or when $\alpha$ increases beyond a critical value ($\alpha_c$).
In some ways, this is similar to the usual BCS-BEC crossover problem, 
where the quasiparticle/quasihole excitation spectrum changes behavior as a function of increasing
the scattering parameter $1/(k_F a_s)$ at $\mu = 0$, i.e. its minimum is located at a 
finite (zero) momenta when $\mu > 0$ ($\mu < 0$). 

For the ERD- and FA-type SOC, we obtain $\mu = \epsilon_F - m\alpha^2/2$, and 
thus setting $\mu = 0$ gives $\alpha_c = k_F/m$.
For the PRD-type SOC, we obtain 
$
m\sqrt{m} \sqrt{2\mu}(2\mu+ 9m \alpha^2/2) = k_F^3
$ 
for $\mu \ge 0$ leading to $\mu \approx \epsilon_F - 3m\alpha^2/2$ in the weak SOC limit
when $m\alpha^2 \ll \epsilon_F$, and also obtain 
$
\mu = 2k_F^3/(3\pi m^2 \alpha) - m\alpha^2/2
$ 
for $\mu \le 0$. Thus, setting $\mu = 0$ gives 
$
\alpha_c = [4/(3\pi)]^{1/3} k_F/m \approx 0.75 k_F/m.
$
Lastly, for the FS-type SOC, we obtain
$
2m\sqrt{m}(\mu + 2m\alpha^2)\sqrt{2\mu + m\alpha^2} = k_F^3,
$
leading to 
$
\mu = -3m\alpha^2/2 + m^3\alpha^4/[4(2m^6 \alpha^6 + k_F^6 + k_F^3\sqrt{4m^6\alpha^6 + k_F^6})]^{1/3}
+ (m^6 \alpha^6 + k_F^6/2 + k_F^3\sqrt{4m^6\alpha^6 + k_F^6}/2)^{1/3}/(2m).
$
Thus, again setting $\mu = 0$ gives
$
\alpha_c = (1/4)^{1/3} k_F/m \approx 0.63 k_F/m.
$

We emphasize that the aforementioned change in Fermi surface topology for the
noninteracting balanced Fermi gases is not a quantum phase transition 
but a smooth crossover~\cite{iskin1, shenoy2}. 
Adding the interactions does not effect this picture much~\cite{shenoy2, zhai, hui, iskin1, han, iskin2}, 
and the crossover in the Fermi surface topology again occurs at $\mu = 0$.
However, this is no longer the case for the interacting imbalanced 
Fermi gases~\cite{gong, iskin1, wyi, iskin2}, and next we show how this crossover 
picture becomes a quantum phase transition in the presence of a population 
imbalance within the self-consistent mean-field theory.

\subsection{Mean-Field Theory}
\label{sec:mft}

In the absence of a SOC and at low temperatures, it is well-established that the 
mean-field theory is sufficient to describe the Fermi gases both in the BCS and 
the BEC limits, and that this theory also captures qualitatively 
the correct physics in the entire BCS-BEC evolution~\cite{review, rmp}. 
Hoping that the mean-field formalism remains sufficient in the presence of a SOC, 
here we analyze the resultant ground-state phase diagrams.

For this purpose, we use the mean-field Hamiltonian
\begin{widetext}
\begin{align}
\label{eqn:ham}
H = \frac{1}{2} \sum_{\mathbf{k}} \psi_\mathbf{k}^\dagger 
 \left( \begin{array}{cccc}
\xi_{\mathbf{k},\uparrow} + S_{k_z} & S_\mathbf{k_\perp} & 0 & \Delta \\
S_\mathbf{k_\perp}^* & \xi_{\mathbf{k},\downarrow} - S_{k_z}& -\Delta & 0  \\
0 & -\Delta^* & -\xi_{\mathbf{k},\uparrow} + S_{k_z} & S_\mathbf{k_\perp}^* \\
\Delta^* & 0 &  S_\mathbf{k_\perp} & -\xi_{\mathbf{k},\downarrow} - S_{k_z}
\end{array} \right)
\psi_\mathbf{k} 
+ \sum_{\mathbf{k}} \xi_{\mathbf{k},+} + \frac{|\Delta|^2}{g},
\end{align}
\end{widetext}
where
$
\psi_{\mathbf{k}}^\dagger = 
[a_{\mathbf{k},\uparrow}^\dagger, a_{\mathbf{k},\downarrow}^\dagger,  a_{\mathbf{-k},\uparrow}, a_{\mathbf{-k},\downarrow}]
$
denotes the fermionic operators collectively,
$
\xi_{\mathbf{k},\sigma} = \epsilon_{\mathbf{k},\sigma} - \mu_\sigma,
$
and 
$
\Delta = g\langle a_{\mathbf{k},\uparrow} a_{-\mathbf{k},\downarrow} \rangle
$
is the mean-field order parameter where $\langle \cdots \rangle$
is the thermal average.
The mean-field thermodynamic potential can be written as~\cite{kubasiak}
\begin{align}
\label{eqn:Omega}
\Omega = \frac{T}{2} \sum_{\mathbf{k},\lambda} \ln \left( \frac{1 + X_{\mathbf{k},\lambda}}{2} \right)
+ \sum_{\mathbf{k}} \xi_{\mathbf{k},+} + \frac{|\Delta|^2}{g},
\end{align}
where $T$ is the temperature, $\lambda = \lbrace 1,2,3,4 \rbrace$ labels the
quasiparticle/quasihole excitation energies $E_{\mathbf{k},\lambda}$,
$
X_{\mathbf{k},\lambda} = \tanh[E_{\mathbf{k},\lambda}/(2T)],
$ 
and
$
\xi_{\mathbf{k},s} = \epsilon_{\mathbf{k},s} - \mu_s
$
where $\mu_s = (\mu_\uparrow + s \mu_\downarrow)/2$.
Here, the quasiparticle/quasihole excitation energies $E_{\mathbf{k},\lambda}$ are 
determined by the eigenvalues of the Hamiltonian matrix given in Eq.~(\ref{eqn:ham}).
Following the usual procedure, i.e. $\partial \Omega / \partial |\Delta| = 0$ for the order 
parameter and $N_\uparrow + s N_\downarrow = - \partial \Omega / \partial \mu_s$ for the 
number equations, we obtain the self-consistency equations
\begin{align}
\label{eqn:gap}
\frac{2|\Delta|}{g} &= \frac{1}{4} \sum_{\mathbf{k},\lambda} \frac{\partial E_{\mathbf{k},\lambda}}{\partial |\Delta|} \left( X_{\mathbf{k},\lambda} - 1 \right), \\
\label{eqn:ntot}
N_\uparrow \pm N_\downarrow &= \frac{1}{4} \sum_{\mathbf{k},\lambda} \left[ \frac{1 \pm 1}{2} + \frac{\partial E_{\mathbf{k},\lambda}}{\partial \mu_\pm} \left( X_{\mathbf{k},\lambda} -1 \right) \right].
\end{align}
These equations are the generalization of the mean-field order parameter
and number equations to the case of an anisotropic SOC, and they are consistent with
the known results in the appropriate limits~\cite{shenoy2, gong, zhai, hui, iskin1, han, iskin2} 
(see Sec.~\ref{sec:qpe}).

As usual, we checked the stability of the mean-field solutions for the uniform 
superfluid phase using the curvature criterion~\cite{iskin1, iskin2}, 
which says that the curvature of $\Omega$ with respect to $|\Delta|$, i.e.
\begin{align}
\frac{\partial^2 \Omega}{\partial |\Delta|^2} = \frac{1}{4} \sum_{\mathbf{k},\lambda}  
&\left[
 \left( \frac{1}{|\Delta|} \frac{\partial E_{\mathbf{k},\lambda}}{\partial |\Delta|} 
-\frac{\partial^2 E_{\mathbf{k},\lambda}}{\partial |\Delta|^2} \right)
 \left( X_{\mathbf{k},\lambda} - 1 \right) \right. \nonumber \\
&\left.- \frac{1}{2T} \left( \frac{\partial E_{\mathbf{k},\lambda}}{\partial |\Delta|} \right)^2 Y_{\mathbf{k},\lambda} 
\right],
\end{align}
needs to be positive, where 
$
Y_{\mathbf{k},\lambda} = \textrm{sech}^2[E_{\mathbf{k},\lambda}/(2T)].
$ 
When the curvature $\partial^2 \Omega / \partial |\Delta|^2$ is negative, 
the uniform mean-field solution does not correspond to a minimum of $\Omega$, and a 
nonuniform superfluid phase, e.g.  a phase separation, is favored. It is known
that the curvature criterion correctly discards the unstable solutions, but metastable 
solutions may still survive. This may cause minor quantitative changes in the 
first order phase transition boundaries~\cite{wyi}.

\subsection{Quasiparticle/Quasihole Excitations}
\label{sec:qpe}

While the eigenvalues of the Hamiltonian matrix given in Eq.~(\ref{eqn:ham}) 
do not acquire a simple analytic form for a general SOC, next we discuss three limits
where the quasiparticle/quasihole excitation energies simplify considerably, 
allowing for further analytical investigation.

\subsubsection{Balanced Fermi gases with $S_\mathbf{k_\perp} \ne 0$ and $S_{k_z} \ne 0$}
\label{sec:bal}

First of all, for balanced ($\xi_{\mathbf{k}, \uparrow} = \xi_{\mathbf{k}, \downarrow}$) Fermi gases, 
we obtain~\cite{rashba, shenoy2, zhai, hui, iskin1, han, chen}
\begin{align}
\label{eqn:Ekbal}
E_{\mathbf{k},\lambda} = s_\lambda \sqrt{\left( \xi_{\mathbf{k},+} + p_\lambda \sqrt{|S_\mathbf{k_\perp}|^2 + S_{k_z}^2} \right)^2 + |\Delta|^2},
\end{align}
where $s_1 = s_2 = +$ and $s_3 = s_4 = -$, and $p_\lambda = -(-1)^\lambda$, i.e. 
$p_1 = p_3 = +$ and $p_2 = p_4 = -$. Since all $E_{\mathbf{k},\lambda}$ 
have no zeros and are always gapped for all parameters in $\mathbf{k}$ space, 
we expect the BCS-BEC evolution to be a smooth crossover for 
balanced Fermi gases even in the presence of a SOC~\cite{shenoy2, zhai, hui, iskin1, han}. 
This is similar to what happens in the usual BCS-BEC crossover picture of a Fermi 
gas with no SOC, where the quasiparticle/quasihole excitation energies are also gapped. 
Therefore, the ground state of balanced Fermi gases with SOC is a topologically 
trivial superfluid~\cite{iskin1, wyi, iskin2}.

In this case, the order parameter equation reduces to
$
1/g  = (1/2) \sum_{\mathbf{k},s} X_{\mathbf{k},s}/(2E_{\mathbf{k},s})
$
and the number equation reduces to
$
N = (1/2) \sum_{\mathbf{k},s} [ 1 - (\xi_{\mathbf{k},s} / E_{\mathbf{k},s}) X_{\mathbf{k},s} ],
$
where $E_{\mathbf{k},+(-)} = E_{\mathbf{k},1(2)}$~\cite{rashba, shenoy2, zhai, hui, han, chen}.
These equations are analytically tractable in the strong-coupling limit when $\mu < 0$ 
and $|\mu| \gg |\Delta|$, for which they are approximately given by
$
1/g \approx (1/2) \sum_{\mathbf{k},s} 1/[2(\varepsilon_{\mathbf{k},s}-\mu)]
$
and
$
N \approx (|\Delta|^2/2) \sum_{\mathbf{k},s} 1/[2(\varepsilon_{\mathbf{k},s}-\mu)^2],
$
respectively, where $\varepsilon_{\mathbf{k},s}$ is given in Eq.~(\ref{eqn:ehel}). 
Comparing the order parameter equation with the bound-state one immediately leads to
$
\mu = (\epsilon_b - m\alpha^2)/2
$
for all four types of SOC, but the number equation leads to
$
|\Delta|^2 = 16\sqrt{\epsilon_F^3 (2|\mu| - m\alpha^2)}/(3\pi) 
$
for the ERD- and FA-,
$
|\Delta|^2 = 2\sqrt{2} k_F^3 (2|\mu|-m\alpha^2) / (3\pi m\sqrt{m|\mu|}) 
$
for the PRD-, and
$
|\Delta|^2 = 2 k_F^3 (2|\mu|-m\alpha^2)^{3/2} / (3\pi m\sqrt{m}|\mu|)
$
for the FS-type SOC.  We note that $|\Delta|$ is independent of $\alpha$ only in
the ERD- and FA-type SOC, which is consistent with the recent numerical 
findings~\cite{shenoy2, han}. This is not surprising since the SOC term can be 
eliminated by a momentum shift in the $x$ and $z$ directions, respectively,
in the ERD- and FA-type SOC, which also leads to a shift in the chemical potential
$
\mu(\alpha) = \mu(0) - m\alpha^2/2.
$
Since Eq.~(\ref{eqn:Ekbal}) depends only on the total magnitude of the SOC,
i.e. $\sqrt{|S_\mathbf{k_\perp}|^2 + S_{k_z}^2}$, the ERD-, FA-, PRD- and FS-type 
SOC differ in their Jacobians of the $\mathbf{k}$-space integrals.

\subsubsection{$S_\mathbf{k_\perp} \to 0$ and $S_{k_z} \ne 0$}
\label{sec:Sz}

When $S_\mathbf{k_\perp} \to 0$ and $S_{k_z} \ne 0$, i.e. the FA-type SOC, we obtain
\begin{align}
E_{\mathbf{k},\lambda} &= s_\lambda \sqrt{\left( \xi_{\mathbf{k},+} + p_\lambda S_{k_z} \right)^2+|\Delta|^2} + p_\lambda \xi_{\mathbf{k},-},
\end{align}
which again can be gapless in some $\mathbf{k}$ space regions. 
The zeros of $E_{\mathbf{k},\lambda}$ can be found by imposing the condition
$
E_{\mathbf{k},1(2)} E_{\mathbf{k},3(4)} = \xi_{\mathbf{k},-}^2 - (\xi_{\mathbf{k},+} + p_\lambda S_{k_z})^2 - |\Delta|^2 = 0
$
indicating that the zeros occur at real $(k_\perp = \sqrt{k_x^2 + k_y^2}, k_z)$ momenta such that
$
k_{\perp,s}^2 = 2 m \mu_+ - k_{z,s}^2 + 2m S_{k_{z,s}} + 2m s\sqrt{\mu_-^2 - |\Delta|^2}.
$
Since $k_{\perp,s} \ge 0$, setting $k_{\perp,s} = 0$ above leads to
$
k_{z,s} = m\alpha \pm \sqrt{2m(\mu_+ + m\alpha^2/2) + 2ms \sqrt{\mu_-^2 - |\Delta|^2}},
$
provided that $|\Delta| < |\mu_-|$ and 
$
|\Delta|^2 < - (\mu_\uparrow + m\alpha^2/2) (\mu_\downarrow + m\alpha^2/2).
$
This analysis shows that the conditions $|\Delta| = |\mu_-|$ and 
$|\Delta|^2 = - (\mu_\uparrow + m\alpha^2/2) (\mu_\downarrow + m\alpha^2/2)$
determine the phase boundaries between the SF, GSF(I) and GSF(II) 
regions (see Sec.~\ref{sec:phase}), 
and that these three phases meet at a tri-critical point determined by $\mu_+ = - m \alpha^2/2$.

In this case, the derivatives of the quasiparticle/quasihole energies are given by
$
\partial E_{\mathbf{k},\lambda} / \partial |\Delta| = s_\lambda |\Delta|/B_{\mathbf{k},p_\lambda}
$
for the order parameter,
$
\partial E_{\mathbf{k},\lambda} / \partial \mu_+ =  -s_\lambda(\xi_{\mathbf{k},+} + p_\lambda S_{k_z})/B_{\mathbf{k},p_\lambda}
$
for the average chemical potential, and
$
\partial E_{\mathbf{k},\lambda} / \partial \mu_- = - p_\lambda
$
for the half of the chemical potential difference. 
Here, $B_{\mathbf{k},p_\lambda} = \sqrt{(\xi_{\mathbf{k},+} + p_\lambda S_{k_z})^2 + |\Delta|^2}$.
We again note for this case that the order parameter equation reduces to
$
2|\Delta|/g = (1/4) \sum_{\mathbf{k},\lambda} (\partial E_{\mathbf{k},\lambda} / \partial |\Delta|) X_{\mathbf{k},\lambda},
$
and the number equations reduce to
$
N_\uparrow \pm N_\downarrow = (1/4) \sum_{\mathbf{k},\lambda} [ (1 \pm 1)/2 + (\partial E_{\mathbf{k},\lambda} / \partial \mu_\pm) X_{\mathbf{k},\lambda} ].
$

Similar to the balanced case discussed in Sec.~\ref{sec:bal}, we note that 
FA-type SOC term again can be eliminated for the population-imbalanced 
Fermi gases, by a momentum shift in the $z$ direction~\cite{massnote}. 
This also leads to a shift in the chemical potentials, i.e. 
$
\mu_\sigma(\alpha) = \mu_\sigma(0) - m\alpha^2/2.
$ 
Therefore, FA-type SOC does not have any observable effect on the phase diagrams,
which is in sharp contrast with the ERD-type SOC as discussed next.

\subsubsection{$S_{k_z} \to 0$ and $S_\mathbf{k_\perp} \ne 0$}
\label{sec:Sperp}

\begin{figure} [htb]
\centerline{\scalebox{0.6}{\includegraphics{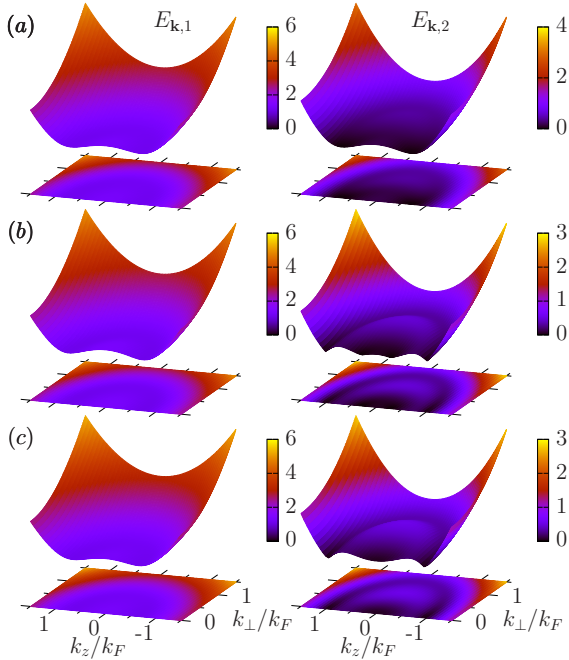}}}
\caption{\label{fig:Ek} (Color online)
The excitation spectra $E_{\mathbf{k}, \lambda}$ of the PRD-type SOC are shown for 
(a) SF phase ($\alpha=0.275 k_F/m$ and $P=0.25$),
(b) GSF(II) phase ($\alpha=0.275 k_F/m$ and $P=0.5$), and
(c) GSF(I) phase ($\alpha=0.35 k_F/m$ and $P=0.5$).
These data correspond to the red cross marks in Fig.~\ref{fig:ai}(b).
}
\end{figure}

On the other hand, when $S_{k_z} \to 0$ and $S_\mathbf{k_\perp} \ne 0$, it is straightforward 
to show that~\cite{kubasiak, gong, iskin1, wyi, iskin2}
\begin{align}
E_{\mathbf{k},\lambda} &= s_\lambda \sqrt{\xi_{\mathbf{k},+}^2+\xi_{\mathbf{k},-}^2+|\Delta|^2+|S_\mathbf{k_\perp}|^2 + 2 p_\lambda A_{\mathbf{k}}},
\end{align}
where 
$
A_{\mathbf{k}} = \sqrt{\xi_{\mathbf{k},-}^2(\xi_{\mathbf{k},+}^2 + |\Delta|^2) + |S_\mathbf{k_\perp}|^2 \xi_{\mathbf{k},+}^2}.
$
Note in this case that $E_{\mathbf{k},\lambda}$ can be gapless at some points/lines in $\mathbf{k}$ 
space. The zeros of $E_{\mathbf{k},\lambda}$ can be found by imposing the condition
$
E_{\mathbf{k},1(3)}^2 E_{\mathbf{k},2(4)}^2 = (\xi_{\mathbf{k},+}^2 - \xi_{\mathbf{k},-}^2 
+ |\Delta|^2 - |S_\mathbf{k_\perp}|^2)^2+4|\Delta|^2 |S_\mathbf{k_\perp}|^2 = 0,
$
indicating that both $|S_\mathbf{k_\perp}| = 0$ and 
$\xi_{\mathbf{k},\uparrow} \xi_{\mathbf{k},\downarrow} + |\Delta|^2 = 0$
needs to be satisfied. 
Therefore, in the PRD-type SOC, the zeros occur when~\cite{iskin1, iskin2} $k_x = k_y = 0$ and at real 
$k_z$ momenta,
$
k_{z,s}^2 = 2m \mu_+ + 2m s \sqrt{\mu_-^2 - |\Delta|^2},
$
provided that $|\Delta| < |\mu_-| $ for $\mu_+ \ge 0$, and 
$|\Delta|^2 < - \mu_\uparrow \mu_\downarrow$ for $\mu_+ < 0$.
Similarly, in the ERD-type SOC, the zeros occur when $k_x = 0$ and at real 
$k_{\rho} = \sqrt{k_y^2 + k_z^2}$ momenta,
$
k_{\rho,s}^2 = 2m \mu_+ + 2m s \sqrt{\mu_-^2 - |\Delta|^2},
$
provided with the same conditions as above.
This analysis shows for both PRD- and ERD-type SOC that the 
conditions $|\Delta| = |\mu_-|$ and 
$|\Delta|^2 = -\mu_\uparrow \mu_\downarrow$ determine the phase 
boundaries between the SF, GSF(I) and GSF(II) regions (see Sec.~\ref{sec:phase}), 
such that these three phases meet at a tri-critical point determined 
by $\mu_+ = 0$~\cite{iskin1, wyi, iskin2}. In Fig.~\ref{fig:Ek}, typical excitation 
spectra $E_{\mathbf{k}, \lambda}$ of the PRD-type SOC are shown for the
SF, GSF(I) and GSF(II) phases, illustrating the $\mathbf{k}$-space topology of 
their gapped/gapless excitations as discussed in Sec.~\ref{sec:phase} in great detail.

In this case, the derivatives of the quasiparticle/quasihole energies are given by
$
\partial E_{\mathbf{k},\lambda} / \partial |\Delta| = (1 + p_\lambda \xi_{\mathbf{k},-}^2/A_{\mathbf{k}} ) |\Delta| / E_{\mathbf{k},\lambda}
$
for the order parameter,
$
\partial E_{\mathbf{k},\lambda} / \partial \mu_+ = - [1 + p_\lambda ( \xi_{\mathbf{k},-}^2+|S_\mathbf{k_\perp}|^2 )/A_{\mathbf{k}} ] \xi_{\mathbf{k},-} / E_{\mathbf{k},\lambda}
$
for the average chemical potential, and
$
\partial E_{\mathbf{k},\lambda} / \partial \mu_- = - [1 + p_\lambda ( \xi_{\mathbf{k},+}^2+|\Delta|^2 )/A_{\mathbf{k}} ] \xi_{\mathbf{k},-} / E_{\mathbf{k},\lambda}
$
for the half of the chemical potential difference.
Therefore, the order parameter equation reduces to
$
2|\Delta|/g = (1/2) \sum_{\mathbf{k},s} (\partial E_{\mathbf{k},s} / \partial |\Delta|) X_{\mathbf{k},s},
$
and the number equations reduce to
$
N_\uparrow \pm N_\downarrow = (1/2) \sum_{\mathbf{k},s} [ (1 \pm 1)/2 + (\partial E_{\mathbf{k},s} / \partial \mu_\pm) X_{\mathbf{k},s} ],
$
where $E_{\mathbf{k},+(-)} = E_{\mathbf{k},1(2)}$~\cite{iskin1, iskin2}.

\begin{table}
\label{tab:top}
\begin{tabular}{|c|c|c|c|c|}
\hline
           	   & no-SOC	& Balanced & ERD & PRD \\
\hline
   SF             & gapped	&   gapped &    gapped   & gapped \\
\hline
   GSF(I)      & 1 surface	&    N/A    &    1 ring  & 2 points \\
\hline
   GSF(II)     & 2 surfaces &    N/A    &    2 rings   & 4 points \\
\hline
\end{tabular}
\caption{The topological classification of uniform superfluid phases are summarized, 
depending on the number of zero energy quasiparticle/quasihole excitation energy 
surfaces, rings or points in $\mathbf{k}$-space.} 
\end{table}

\subsection{Ground-State Phase Diagrams}
\label{sec:phase}

\begin{figure} [htb]
\centerline{\scalebox{0.5}{\includegraphics{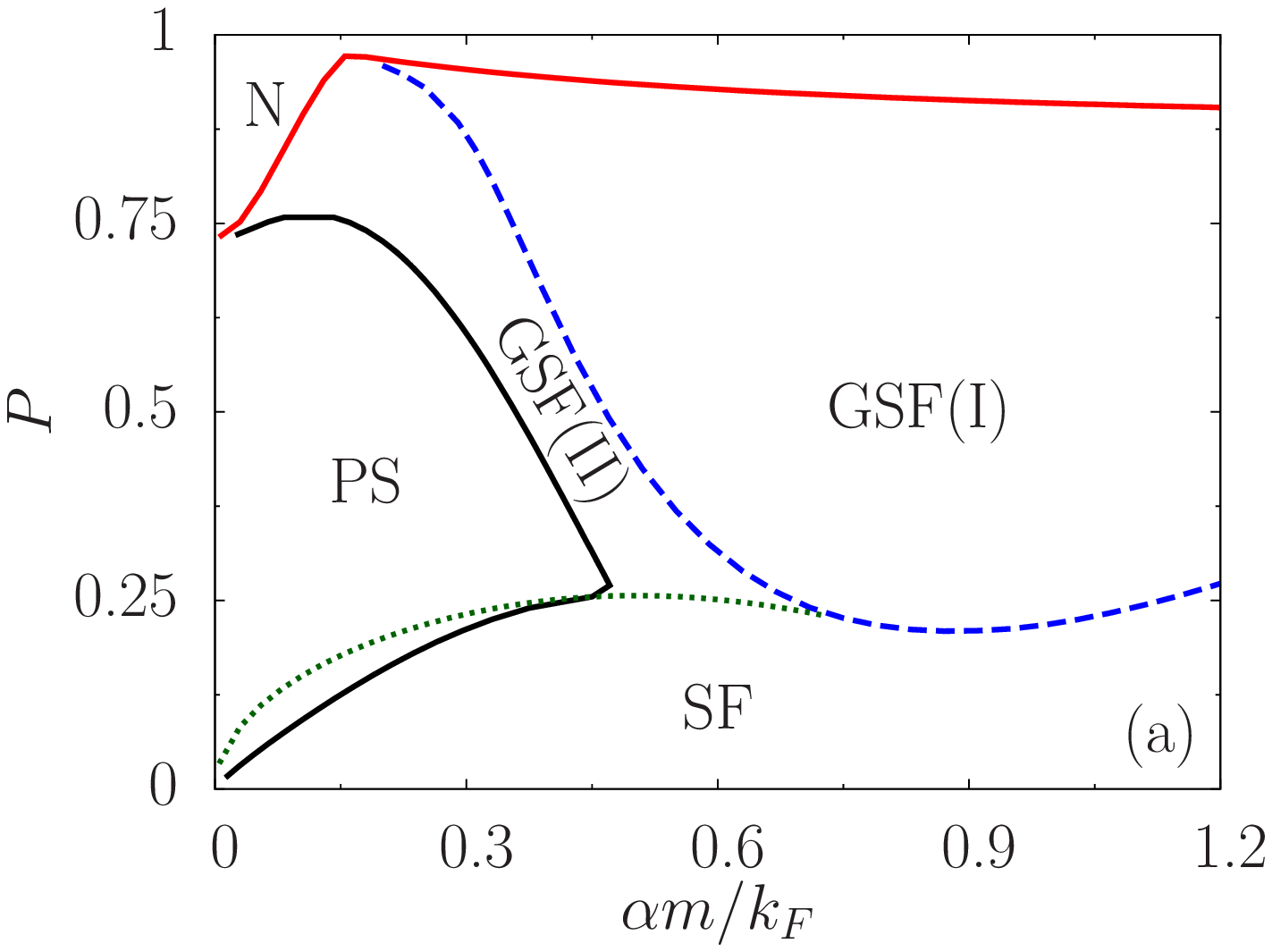}}}
\centerline{\scalebox{0.5}{\includegraphics{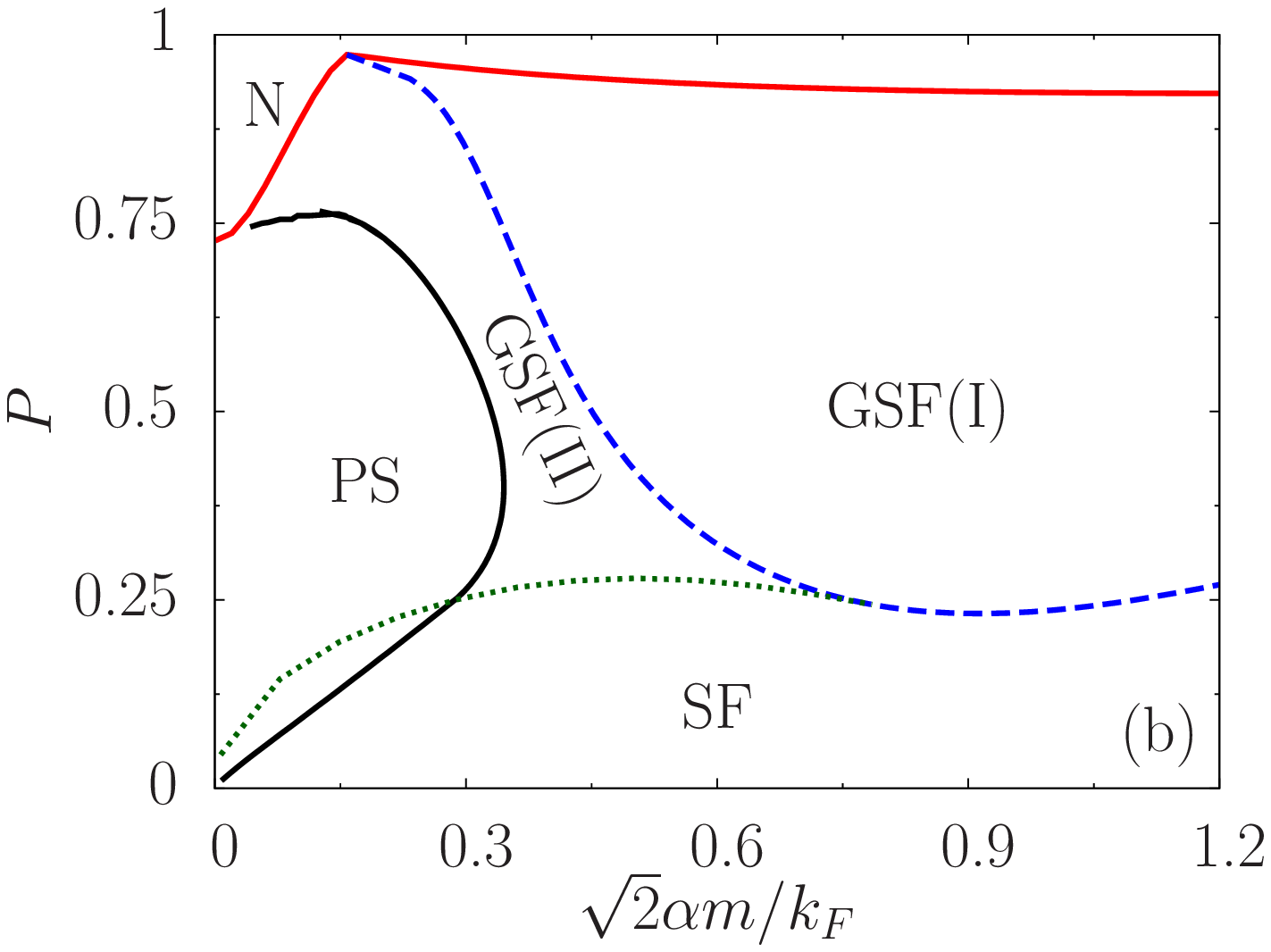}}}
\caption{\label{fig:ai} (Color online)
The ground-state phase diagrams of (a) ERD- and (b) PRD-type SOC 
are shown as a function of $P = (N_\uparrow - N_\downarrow)/N$ 
and $\alpha$ at unitarity, i.e. when $1/(k_F a_s) = 0$. 
The phase labels are described in the text (see Sec.~\ref{sec:phase}).
Note that the regions bounded by the solid black lines show instability toward  
a non-uniform superfluid phase (PS), and the dashed and dotted lines shown 
within these regions are solely for illustration purposes.
In (b) the red crosses mark a point from each phase for which the corresponding 
excitation spectra and momentum distributions are shown in Figs.~\ref{fig:Ek} 
and~\ref{fig:nk}, respectively.
}
\end{figure}

There are three phases in the phase diagrams~\cite{iskin1, wyi, iskin2}. 
While the normal (N) phase is characterized by $\Delta = 0$, the uniform superfluid
and nonuniform superfluid, e.g. phase separation (PS), are 
characterized by $\partial^2\Omega / \partial |\Delta|^2 > 0$ and 
$\partial^2\Omega / \partial |\Delta|^2 < 0$, respectively, when $\Delta \ne 0$. 
Furthermore, in addition to the topologically trivial gapped superfluid (SF) phase, 
the gapless superfluid (GSF) phase can also be distinguished by the
momentum-space topology of its excitations. Depending on the number of zeros of 
$E_{\mathbf{k},\lambda}$ (zero energy regions in $\mathbf{k}$ space), there are 
two topologically distinct gapless phases. For the ERD-type SOC, 
we have GSF(I) where $E_{\mathbf{k},\lambda}$ has one, and GSF(II) where 
$E_{\mathbf{k},\lambda}$ has two zero energy rings in $\mathbf{k}$ space.  
Similarly, for the PRD-type SOC, we have GSF(I) 
where $E_{\mathbf{k},\lambda}$ has two, and GSF(II) where $E_{\mathbf{k},\lambda}$ 
has four zero energy points in $\mathbf{k}$ space. The topological classification 
of uniform superfluid phases are summarized in Table~I.
In Fig.~\ref{fig:Ek}, we show the excitation spectra $E_{\mathbf{k}, \lambda}$ of the PRD-type SOC 
for the SF phase in~\ref{fig:Ek}(a), GSF(II) phase in~\ref{fig:Ek}(b), and GSF(I) phase in~\ref{fig:Ek}(c),
illustrating $\mathbf{k}$-space topology of their gapped/gapless excitations.
These data correspond to the points indicated by the red cross marks in Fig.~\ref{fig:ai}(b).

\begin{figure} [htb]
\centerline{\scalebox{0.5}{\includegraphics{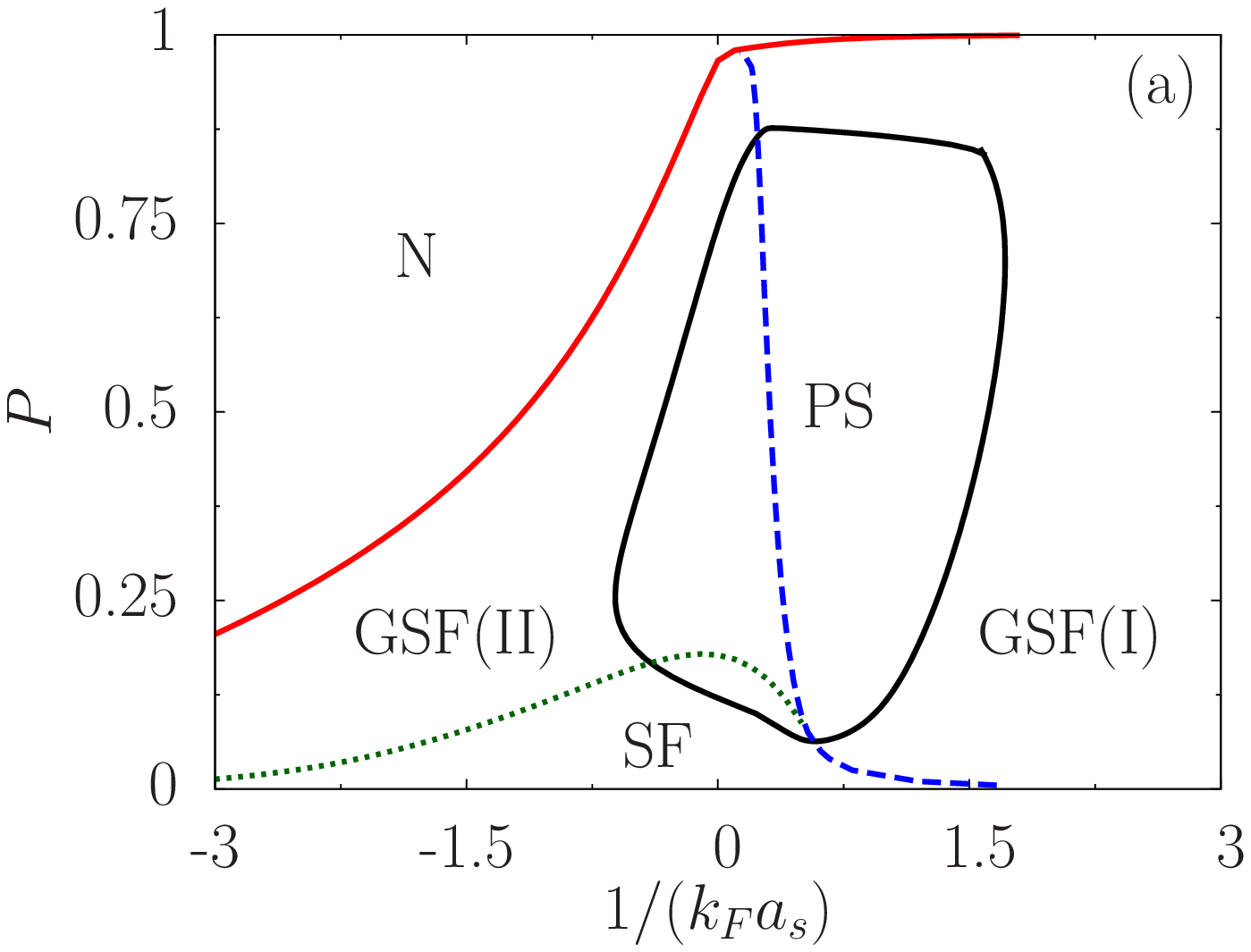}}}
\centerline{\scalebox{0.5}{\includegraphics{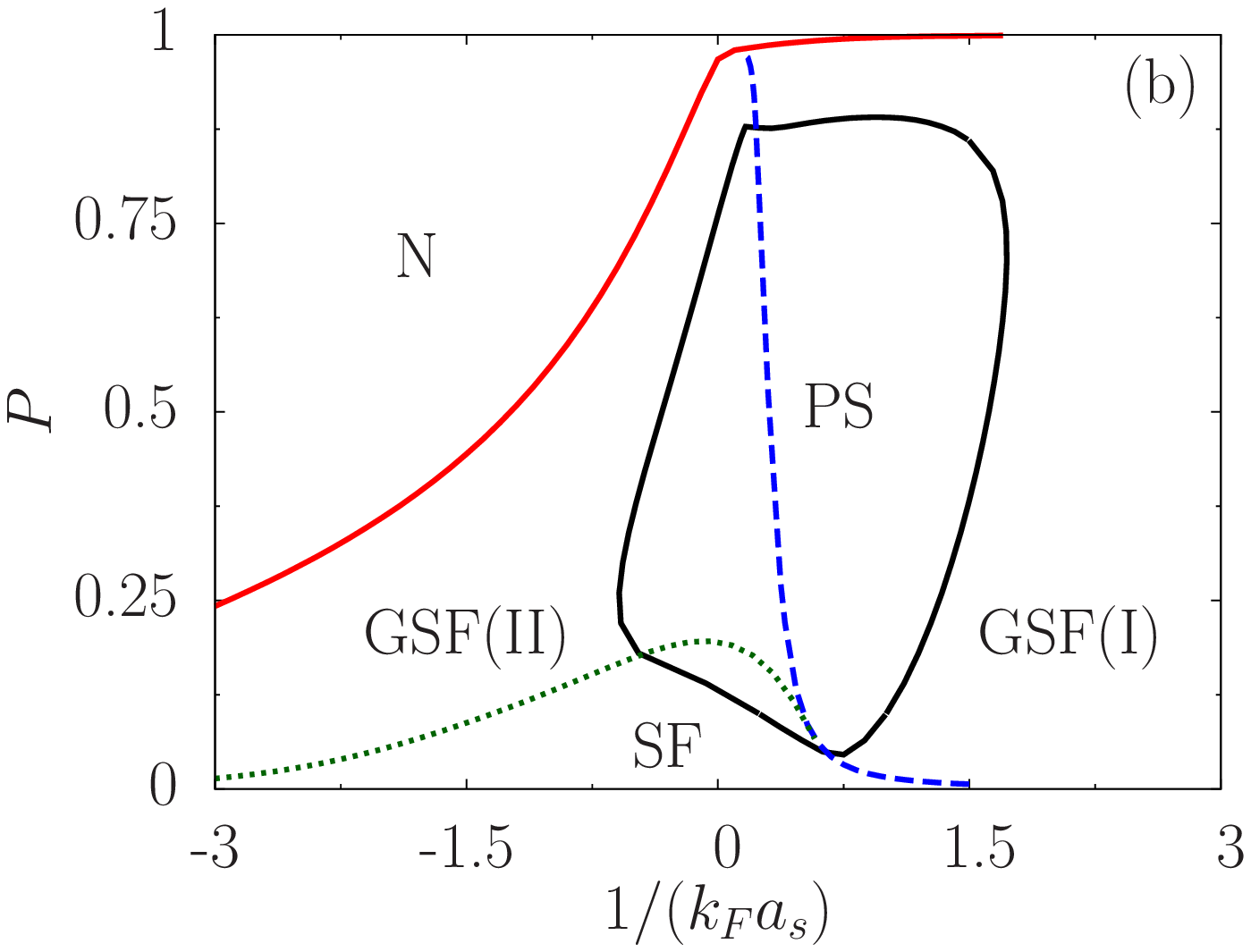}}}
\caption{\label{fig:alpha} (Color online) 
The ground-state phase diagrams of (a) ERD- and (b) PRD-type SOC 
Fermi gases are shown as a function of $P = (N_\uparrow - N_\downarrow)/N$ 
and $1/(k_F a_s)$. Here, we choose $\alpha = 0.15 k_F/m$ in (a) and 
$\alpha = (0.15/\sqrt{2}) k_F/m$ in (b), so that the magnitudes of the SOC
are the same in both figures. 
The phase labels are described in the text (see Sec.~\ref{sec:phase}).
Note that the regions bounded by the solid black lines show instability toward  
a non-uniform superfluid phase (PS), and the dashed and dotted lines shown 
within these regions are solely for illustration purposes.
}
\end{figure}

In Fig.~\ref{fig:alpha}, we show the ground-state phase diagrams of (a) ERD- 
and (b) PRD-type SOC Fermi gases as a function of the population imbalance 
$P = (N_\uparrow - N_\downarrow)/N$ and the scattering parameter $1/(k_F a_s)$.
Here, the dashed blue and dotted green lines correspond to 
$|\Delta|^2 = -\mu_\uparrow \mu_\downarrow$ and $|\Delta| = |\mu_-|$, respectively,
and they mark the SF, GSF(I) and GSF(II) phase boundaries. We note that the regions 
bounded by the solid black lines show instability toward  a non-uniform superfluid 
phase (PS). Since our classification of distinct topological phases applies 
only to the uniform superfluid region, the dashed and dotted lines shown within 
these regions are solely for illustration purposes.

We find that while the ERD-type SOC does not have any observable effect
on the balanced Fermi gases (see Sec.~\ref{sec:bal}), it gives rise to a phase 
diagram with very similar topological structure as that of the PRD-type SOC.
This is not very surprising since the main difference between the ERD- and 
PRD-type SOC is the Jacobians involved in the $\mathbf{k}$-space integrals.
This must be contrasted with the FA-type SOC, which does not have any 
observable effect on the system even in the presence of a population imbalance 
(see Sec.~\ref{sec:Sz}), and therefore, its phase diagram is exactly the same 
as that of the usual population-imbalanced Fermi gases without the SOC.
Since this problem is well-studied in the literature~\cite{review, rmp}, we do 
not discuss it any further.  

Comparing Fig.~\ref{fig:alpha} with the $\alpha \to 0$ limit~\cite{review, rmp}, it 
is clearly seen that both the ERD- and PRD-type SOC are counteracting 
the population imbalance. On one hand, this competition always tends to stabilize 
the GSF phase against the PS, and therefore, at any given $P$, the system 
eventually transitions to a stable SF or GSF by increasing $\alpha$, 
no matter how small $1/(k_F a_s)$ is. 
This is best seen in Fig.~\ref{fig:ai}, where the phase diagrams are shown as 
a function of $P$ and $\alpha$ at unitarity when $1/(k_F a_s) = 0$.
On the other hand, we find that while both the ERD- and PRD-type SOC stabilize 
the GSF phase against the N phase for low $P$ due to increased density of 
states~\cite{shenoy2, zhai}, they destabilize the GSF phase against the N 
phase for high $P$.

In sharp contrast to the $\alpha = 0$ case where only the gapless 
GSF phase can support population imbalance, one of the intriguing effects 
of the ERD- and PRD-type SOC is that both the gapless GSF and gapped 
SF phases can support population imbalance when $\alpha \ne 0$. This
is possible due to the anisotropic nature of the SOC Fermi gases in 
$\mathbf{k}$ space. In fact, it has recently been shown for the mass-imbalanced 
Fermi gases that both the gapped SF and gapless GSF phases can support 
population balance when $\alpha \ne 0$~\cite{iskin2}.  This is again in sharp contrast to 
the $\alpha = 0$ case where only the gapped SF phase can support 
population balance, and it is possible solely due to the anisotropic nature 
of the SOC Fermi gases in $\mathbf{k}$ space.

\begin{figure} [htb]
\centerline{\scalebox{0.65}{\includegraphics{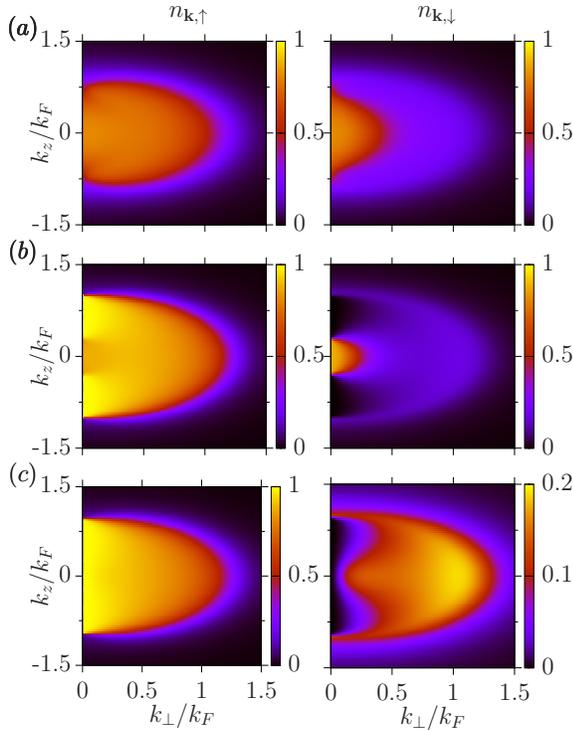}}}
\caption{\label{fig:nk} (Color online)
The momentum distributions $n_{\mathbf{k},\sigma}$ of the PRD-type SOC are shown for 
(a) SF phase,
(b) GSF(II) phase, and
(c) GSF(I) phase.
These data again correspond to the points indicated by red cross marks 
in Fig.~\ref{fig:ai}(b).
}
\end{figure}

The transition from GSF(II) to GSF(I) leads to a change in topology 
in the lowest quasiparticle band, similar to the Lifshitz transition 
in ordinary metals and nodal (non-$s$-wave) superfluids. However, 
the topological transition discussed here is unique, because it involves an $s$-wave 
superfluid, and could be potentially observed for the first time through the 
measurement of the momentum distributions of $\uparrow$
and $\downarrow$ fermions~\cite{iskin1, wyi}. The momentum distributions are 
readily available from Eq.~(\ref{eqn:ntot}), and we illustrate the typical 
$T = 0$ distributions of the SF, GSF(I), and GSF(II) phases in Fig.~\ref{fig:nk}. The 
distributions are anisotropic in $\mathbf{k}$ space, 
which follows from the anisotropic structure of $E_{\mathbf{k},\lambda}$. 
In addition, while the distributions of SF phase do not show sharp features, those 
of GSF(I) and GSF(II) phases are exactly $n_{\mathbf{k},\uparrow} = 1$ and 
$n_{\mathbf{k},\downarrow} = 0$ for $\mathbf{k}$-space regions where 
$k_\perp = 0$ and $k_{z,-} \le |k_z| \le k_{z,+}$. Therefore, a major redistribution 
occurs for the minority component ($n_{\mathbf{k},\downarrow}$) at the topological 
phase transition boundaries. For instance, at the GFS(II) to GSF(I) transition 
boundary, the sharp peak that is present near the origin vanishes abruptly.

Although this topological transition is quantum in its nature, signatures of it should 
still be observed at finite $T$, where the observables are smeared 
out due to thermal effects. While the primary signature of this topological transition is 
seen in the momentum distribution, single-particle spectral function~\cite{rf} 
as well as some thermodynamic quantities such as the atomic compressibility 
would also show an anomaly at the transition boundary. 

All of these results are obtained within the self-consistent mean-field approximation, 
which is known to be reliable for the entire BCS-BEC evolution only near $T = 0$.
Since the fluctuations dominate the physics at finite $T$ towards the molecular 
BEC limit~\cite{carlos}, to emphasize further the effects of finite $T$, next we discuss 
the Gaussian fluctuations near the critical temperature $T_c$, i.e. 
the temperature at which the mean-field order parameter $\Delta$ vanishes.

\section{Gaussian Fluctuations Near $T_c$}
\label{sec:fluct}

One way to go beyond the mean-field (or saddle-point) approximation
and study the Gaussian fluctuations is to use the imaginary-time 
functional integral approach~\cite{carlos, iskinlwave}. 
Using this approach and expanding the order parameter field around 
$\Delta = 0$, one obtains the fluctuation action 
$
S_G = (1/T) \sum_{\mathbf{q},n} L^{-1}(\mathbf{q},\nu_n) |\Lambda(\mathbf{q},\nu_n)|^2,
$
where $L^{-1}(\mathbf{q},\nu_n)$ is the inverse 
fluctuation propagator, $\Lambda(\mathbf{q}, \nu_n)$ is the fluctuation field,
and $\nu_n = 2\pi T n$ is the bosonic Matsubara frequency. It is 
a straightforward task to calculate the propagator
\begin{align}
&L^{-1}(\mathbf{q}, \nu_n) = \frac{1}{g} - \frac{1}{8} \sum_{\mathbf{k},\lambda_o,\lambda_e} 
\frac{X_{\mathbf{k}+\frac{\mathbf{q}}{2}, \lambda_o} - X_{\mathbf{k}-\frac{\mathbf{q}}{2}, \lambda_e}}
{E_{\mathbf{k}+\frac{\mathbf{q}}{2}, \lambda_o} - E_{\mathbf{k}-\frac{\mathbf{q}}{2}, \lambda_e} - i\nu_n} \Big{[}1- \nonumber \\ 
& \left.s_{\lambda_o} p_{\lambda_e} \frac{C_{\mathbf{k}+\frac{\mathbf{q}}{2},+} C_{\mathbf{k}-\frac{\mathbf{q}}{2},-} 
                   - (S_{\mathbf{k_\perp}+\frac{\mathbf{q_\perp}}{2}} S^*_{\mathbf{k_\perp}-\frac{\mathbf{q_\perp}}{2}} + h.c.)/2}
{\sqrt{(C_{\mathbf{k}+\frac{\mathbf{q}}{2},+}^2 + |S_{\mathbf{k_\perp}+\frac{\mathbf{q_\perp}}{2}}|^2) 
             (C_{\mathbf{k}-\frac{\mathbf{q}}{2},-}^2 + |S_{\mathbf{k_\perp}-\frac{\mathbf{q_\perp}}{2}}|^2)}} 
\right],
\end{align}
where $\lambda_o = \lbrace 1,3 \rbrace$ and $\lambda_e = \lbrace 2,4 \rbrace$ sums over
odd and even $\lambda$ values, respectively, $h.c.$ is the Hermitian conjugate,
$
C_{\mathbf{k},s} = \xi_{\mathbf{k},-} + s S_{k_z},
$
and
\begin{align}
E_{\mathbf{k},\lambda} = s_\lambda \sqrt{(\xi_{\mathbf{k},-} + p_\lambda S_{k_z})^2 + |S_\mathbf{k_\perp}|^2} + p_\lambda \xi_{\mathbf{k},+} 
\end{align}
are the quasiparticle/quasihole excitation energies. The calculation of the fourth order fluctuations 
is lengthy and straightforward, but the results are not particularly illuminating. 
Next we use the fluctuation action to study the time-dependent Ginzburg-Landau 
functional near $T_c$.

\subsection{Ginzburg-Landau theory near $T_c$}
\label{sec:GL}

The Ginzburg-Landau theory is used to study the low-frequency and long-wavelength
behavior of the order parameter near $T_c$. For this purpose, first we consider
the static part of the propagator and expand $L^{-1}(\mathbf{q}, 0)$ in powers of 
$q_i$, and then expand $L^{-1}(\mathbf{0},\nu_n) - L^{-1}(\mathbf{0}, 0)$
in powers of $\omega$ after the analytic continuation $i\nu_n \to \omega + i0^+$.

This calculation leads to the time-dependent Ginzburg-Landau equation 
in $\mathbf{k}$ space~\cite{carlos, iskinlwave}, i.e.
$
L^{-1}(\mathbf{q}, \omega) = a(T) + \sum_{i,j} c_{ij} q_i q_j/2 - d \omega.
$
Here, the zeroth order coefficient $L^{-1}(\mathbf{0},0)$ is given by
\begin{align}
a(T) &= \frac{1}{g} - \frac{1}{8} \sum_{\mathbf{k},\lambda_o,\lambda_e} 
\frac{X_{\mathbf{k}, \lambda_o} - X_{\mathbf{k}, \lambda_e}}
{E_{\mathbf{k}, \lambda_o} - E_{\mathbf{k}, \lambda_e}} \nonumber \\
& \times \left[1 - s_{\lambda_o} p_{\lambda_e} \frac{\xi_{\mathbf{k},-}^2 - |S_\mathbf{k_\perp}|^2 - S_{k_z}^2} 
{\sqrt{(C_{\mathbf{k},+}^2 + |S_\mathbf{k_\perp}|^2) (C_{\mathbf{k},-}^2 + |S_\mathbf{k_\perp}|^2)}} 
\right].
\end{align}
The condition $a(T_c) = 0$ is the Thouless criterion, and it 
leads to an equation for $T_c$. We checked for all four types of SOC that 
this criterion is in agreement with the order parameter equation 
after setting $|\Delta| = 0$ in the latter.
The coefficient of the time-dependent term
\begin{align}
d &=  \frac{1}{8} \sum_{\mathbf{k},\lambda_o,\lambda_e} (X_{\mathbf{k}, \lambda_o} - X_{\mathbf{k}, \lambda_e}) \nonumber \\
& \times \left[ 1 - s_{\lambda_o} p_{\lambda_e} \frac{\xi_{\mathbf{k},-}^2 - |S_\mathbf{k_\perp}|^2 - S_{k_z}^2} 
{\sqrt{(C_{\mathbf{k},+}^2 + |S_\mathbf{k_\perp}|^2) (C_{\mathbf{k},-}^2 + |S_\mathbf{k_\perp}|^2)}} 
\right] \nonumber \\ 
& \times
\left[
 \frac{1}{(E_{\mathbf{k}, \lambda_o} - E_{\mathbf{k}, \lambda_e})^2} + \frac{i \delta(E_{\mathbf{k}, \lambda_o} - E_{\mathbf{k}, \lambda_e} - \omega)}{\omega} 
\right]
\end{align}
is a complex number. For balanced Fermi gases, while its imaginary part reflects 
the decay of Cooper pairs into the two-particle continuum for $\mu > 0$, the imaginary part 
vanishes for $\mu < 0$ and the behavior of the fluctuation field is propagating,
reflecting the presence of stable bound states (molecular bosons)~\cite{carlos, iskinlwave}.

The second-order coefficient 
$
\partial^2 L^{-1}(\mathbf{q},0) / \partial q_i \partial q_j
$
evaluated at $\mathbf{q} = 0$ is not illuminating for imbalanced Fermi gases.
However, for balanced Fermi gases with PRD-type SOC, where
$c_{xx} = c_{yy} = c_\perp$, $c_{zz} = c_z$ and $c_{i \ne j} = 0$, the coefficients
are given by
\begin{align}
c_{ij} &= \frac{1}{16m} \sum_{\mathbf{k}, s} \Big\lbrace
\frac{X_{\mathbf{k}, s} Y_{\mathbf{k}, s}}{2m T^2 E_{\mathbf{k,s}}} k_i k_j \left( 1 + s\frac{m\alpha}{k_\perp} \right)^2 \nonumber \\
&- \frac{Y_{\mathbf{k}, s}}{2 T E_{\mathbf{k,s}}} \left[ \delta_{ij} \left(1 + s\frac{m\alpha}{k_\perp} \right)^2 - s\frac{m\alpha}{k_\perp} \frac{k_i k_j}{k_\perp^2} \right] \nonumber \\
&+ \frac{X_{\mathbf{k}, s}}{E_{\mathbf{k,s}}^2} \left[ \delta_{ij}  + s\frac{m\alpha}{k_\perp} \left(\delta_{ij} - \frac{k_i k_j}{k_\perp^2} \right) \right]
\Big\rbrace
\nonumber \\
&+ \frac{1}{8} \sum_{\mathbf{k},s,s'} 
\frac{X_{\mathbf{k}, s} + X_{\mathbf{k}, s'}} {E_{\mathbf{k}, s} + E_{\mathbf{k}, s'}} 
\frac{s s'}{k_\perp^2} \left( \delta_{ij} - \frac{k_i k_j}{k_\perp^2} \right),
\end{align}
where $E_{\mathbf{k}, +(-)} = \pm E_{\mathbf{k}, 1(2)}$. 
One needs to set the explicit $\alpha$ terms and the last $\sum_{\mathbf{k},s,s'}$ term
to zero to extract $c_z$ from this expression. 
For balanced Fermi gases with FS-type SOC, where $c_{ii} = c_0$ and $c_{i \ne j} = 0$, 
one simply needs to replace $k_\perp$ with $k$ to extract $c_0$. 
Similarly, in the case of ERD-type SOC, where $c_{xx} \ne c_{yy} = c_{zz}$ and 
$c_{i \ne j} = 0$, one needs to replace $k_\perp$ with $k_x$ to extract $c_{xx}$, and 
set the explicit $\alpha$ terms and the last $\sum_{\mathbf{k},s,s'}$ term to zero 
to extract $c_{yy} = c_{zz}$.
Finally, in the case of FA-type SOC , where $c_{xx} = c_{yy} \ne c_{zz}$ and 
$c_{i \ne j} = 0$, one needs to replace $k_\perp$ with $k_z$ to extract $c_{zz}$, and 
set the explicit $\alpha$ terms and the last $\sum_{\mathbf{k},s,s'}$ term to zero 
to extract $c_{xx} = c_{yy}$. 

In general, the coefficients $a(T)$, $d$ and $c_{ii}$ need to be 
calculated numerically together with the order parameter and number
equations. However, it can be analytically shown that their asymptotic 
forms recover the usual Ginzburg-Landau equation for BCS 
superfluids in weak coupling and the Gross-Pitaevskii equation 
for a weakly interacting dilute Bose gas in strong coupling~\cite{carlos, iskinlwave}. 
Next we use the latter correspondence to extract the effective mass of the 
Cooper pairs (molecular bosons) and their critical condensation temperature 
in the molecular BEC limit.

\subsection{Molecular BEC limit}
\label{sec:molecular}

For illustration purposes, here we consider only balanced Fermi gases. 
In the molecular BEC limit, when $\mu < 0$ and $|\mu| \gg T_c$ so that 
$E_{\mathbf{k},s}/T \to \infty$, we may set $X_{\mathbf{k}, s} \to 1$ 
and $Y_{\mathbf{k}, s} \to 0$. Therefore, in this limit, 
the time-dependent coefficient simplifies to
$
d = \sum_{\mathbf{k},s} 1/(8 E_{\mathbf{k},s}^2)
$
for all types of SOC. The second-order coefficients simplify to
$
c_\perp = \sum_{\mathbf{k},s} [1 + s m\alpha/(2k_\perp)] / (16m E_{\mathbf{k},s}^2)
+ \sum_{\mathbf{k},s,s'} s s'/ [8k_\perp^2(E_{\mathbf{k}, s} + E_{\mathbf{k}, s'})]
$
along the $(x, y)$ directions and to
$
c_z = \sum_{\mathbf{k},s} 1 / (16m E_{\mathbf{k},s}^2)
$
along the $z$ direction for the PRD-type SOC, and similarly to
$
c_0 = \sum_{\mathbf{k},s} [1 + 2s m\alpha/(3k)] / (16m E_{\mathbf{k},s}^2)
+ \sum_{\mathbf{k},s,s'} s s'/ [6k^2(E_{\mathbf{k}, s} + E_{\mathbf{k}, s'})]
$
along all $(x, y, z)$ directions for the FS-type SOC.
In the case of ERD- and FA-type SOC, we note that since $(\delta_{ij} - k_i k_j/k_x^2) = 0$ 
for $i = j = x$ and $(\delta_{ij} - k_i k_j/k_z^2) = 0$ for $i = j = z$, respectively, 
the diagonal coefficients all become equal in the molecular limit, 
i.e. $c_0 = c_{ii}$, and it is
$
c_0 = \sum_{\mathbf{k},s} 1 / (16m E_{\mathbf{k},s}^2).
$

These $\mathbf{k}$-space sums are analytically tractable for all four types of SOC 
that we consider in this paper. For instance, for the ERD- or FA type SOC, we obtain 
$
d = 2m c_0 = m \sqrt{m} V / (8\pi\sqrt{2|\mu|-m\alpha^2}).
$
However, for the PRD-type SOC, we obtain
$
d = m \sqrt{2m|\mu|} V / [8\pi(2|\mu|-m\alpha^2)]
$
for the time-dependent, and
$
c_\perp= \sqrt{2m} V (4|\mu|-m\alpha^2)/[64\pi\sqrt{|\mu|}(2|\mu|-m\alpha^2)]
-\sqrt{2m}/[64\pi \sqrt{|\mu|}] \ln[(2|\mu| - m\alpha^2)/(2|\mu|)]
$
along the $(x, y)$ directions and
$
c_z = m^2 \sqrt{2m|\mu|} V / [4\pi(2|\mu|-m\alpha^2)]
$
along the $z$ direction for the second-order coefficients. 
Similarly, for the FS-type SOC, we obtain
$
d = m \sqrt{m}|\mu| V / [4\pi(2|\mu|-m\alpha^2)^{3/2}]
$
for the time-dependent and
$
c_0 = \sqrt{m} (7|\mu| - 3m\alpha^2) V / [24\pi(2|\mu|-m\alpha^2)^{3/2}]
-\sqrt{m}/(12\pi\sqrt{2|\mu|})
$
for the second-order coefficients. Next we extract the effective Gross-Pitaevskii 
parameters using the asymptotic forms of the Ginzburg-Landau coefficients given above.

\subsubsection{Effective Molecular Mass}
\label{sec:mB}

We recall that, since the Ginzburg-Landau theory derived above reduces to the 
Gross-Pitaevskii theory of a weakly-interacting molecular Bose gas after the rescaling 
$
\Psi(\mathbf{q}, \omega) = \sqrt{d} \Lambda(\mathbf{q}, \omega),
$ 
the effective mass of the Cooper pairs (molecular bosons) along the $i$th direction is 
simply given by $m_{B,i} = d/c_{ii}$~\cite{carlos, iskinlwave}. In the absence of a
SOC, this gives $m_B = 2m$ for all $(x, y, z)$ directions.

\begin{figure} [htb]
\centerline{\scalebox{0.5}{\includegraphics{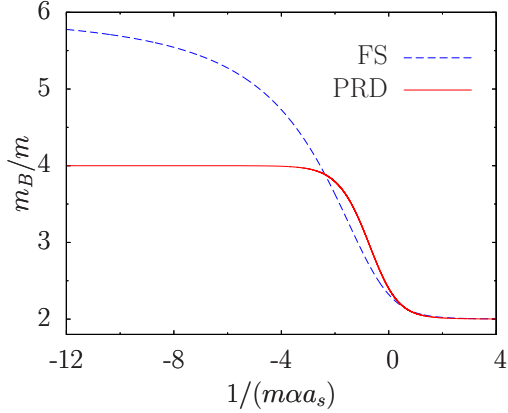}}}
\caption{\label{fig:mB} (Color online)
The effective mass of the Cooper pairs are shown in the molecular BEC limit 
as a function $1/(m \alpha a_s)$ for the PRD- and FS-type SOC.
}
\end{figure}

Using the asymptotic forms of $d$ and $c_0$, and setting $|\mu| = (|\epsilon_b| + m\alpha^2)/2$, 
we find that the mass of the molecular bosons is $m_{B, x} = m_{B, y} = m_{B, z} = m_B = 2m$
for the ERD- and FA-type SOC.
However, using the asymptotic forms of $d$, $c_\perp$ and $c_z$, and setting 
$|\mu| = (|\epsilon_b| + m\alpha^2)/2$, we find that the mass of the molecular 
bosons $m_{B, x} = m_{B, y} = m_{B, \perp}$ is
\begin{align}
\label{eqn:mBperp}
\frac{2m}{m_{B,\perp}} = \frac{2|\epsilon_b| + m\alpha^2}{2|\epsilon_b| + 2m\alpha^2} 
- \frac{|\epsilon_b|}{2|\epsilon_b| + 2m\alpha^2} \ln\left(\frac{|\epsilon_b|}{|\epsilon_b| + m\alpha^2} \right),
\end{align}
and $m_{B,z} = 2m$ along the $z$ direction for the PRD-type SOC.
This expression is in complete agreement with the recent findings~\cite{zhai, hui}, 
showing that $m_{B,\perp}$ decreases monotonically with increasing $1/(m \alpha a_s)$
as plotted in Fig.~\ref{fig:mB}. Equation~(\ref{eqn:mBperp}) gives $m_B = 4m$ 
when $1/(m \alpha a_s) \to -\infty$, $m_B = 2m$ when $1/(m \alpha a_s) \to \infty$, 
and $m_B \approx 2.40m$ at unitarity when $1/(m \alpha a_s) \to 0$.
On the other hand, we find that the mass of the molecular bosons 
$m_{B, x} = m_{B, y} = m_{B, z} = m_B$ is
\begin{align}
\label{eqn:mB}
\frac{2m}{m_B} = \frac{7}{3} - \frac{4}{3}\left(\frac{|\epsilon_b|}{|\epsilon_b| + m\alpha^2} \right)^{3/2} - \frac{2m\alpha^2}{|\epsilon_b| + m\alpha^2}
\end{align}
for the FS-type SOC, which is also a monotonically decreasing function of $1/(m \alpha a_s)$,
as shown in Fig.~\ref{fig:mB}. Equation~(\ref{eqn:mB}) gives $m_B = 6m$ when 
$1/(m \alpha a_s) \to -\infty$, $m_B = 2m$ when $1/(m \alpha a_s) \to \infty$, 
and $m_B  = 3\sqrt{2}m/(2\sqrt{2} -1) \approx 2.32m$ at unitarity.
Having calculated the effective mass of the Cooper pairs, we are ready to calculate their
critical BEC temperature.

\subsubsection{Critical BEC Temperature}
\label{sec:TBEC}

The $\alpha$ dependence of the Cooper pair mass in the molecular BEC limit has a 
dramatic effect on the finite $T$ phase diagram of the system.
For atomic Bose gases, $T_{BEC}$ is determined from the number equation,
$
N_B = \sum_{\mathbf{k}}1/(e^{\epsilon_{\mathbf{k},B}/T_{BEC}} - 1),
$
where $N_B$ is the number and
$
\epsilon_{\mathbf{k},B} = \sum_{i=\lbrace x,y,z \rbrace} k_i^2/(2m_{B,i})
$ 
is the kinetic energy of atomic bosons, with $m_{B,i}$ their effective mass along the 
$i$th direction. This leads to
$
T_{BEC} = 2\pi [n_B/(\sqrt{\Pi_i m_{B,i}} \zeta(3/2))]^{2/3}
$
in three dimensions, where $n_B = N_B/V$ is the density of bosons and $\zeta(x)$ is 
the Riemann zeta function with $\zeta(3/2) \approx 2.61$. Setting $n_B = n/2$,
where $n = N/V = k_F^3/(3\pi^2)$ is the total density of fermions, we obtain 
\begin{align}
T_{BEC} \approx 0.218 \frac{2m}{\left( \Pi_{i=\lbrace x,y,z \rbrace} m_{B,i} \right)^{1/3}} \epsilon_F
\end{align}
in three dimensions, where $\epsilon_F = k_F^2/(2m)$ is the Fermi energy and
$m_{B,i}$ is the mass of the molecular bosons along the $i$th direction.
In the absence of a SOC, this gives $T_{BEC} \approx 0.218 \epsilon_F$~\cite{carlos}.

\begin{figure} [htb]
\centerline{\scalebox{0.5}{\includegraphics{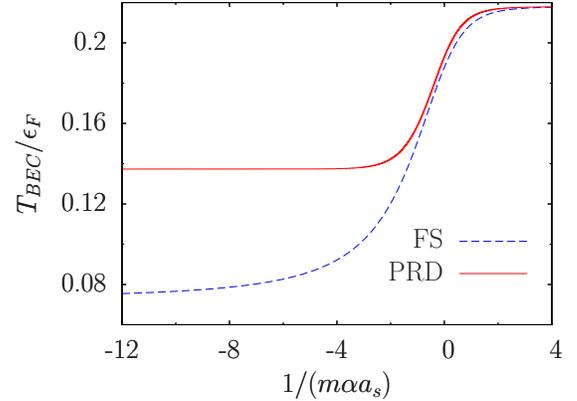}}}
\caption{\label{fig:TBEC} (Color online)
The critical BEC temperature of the Cooper pairs (molecular bosons) are 
shown as a function $1/(m \alpha a_s)$ for the PRD- and FS-type SOC.
}
\end{figure}

Since we expect the Cooper pairs to progressively become weakly-repulsive 
in the molecular BEC limit, and using the effective mass of the Cooper pairs 
found in Sec.~\ref{sec:mB}, we find
$
T_{BEC} \approx 0.218 \epsilon_F
$
for the ERD- and FA-type SOC,
$
T_{BEC} \approx 0.218 (2m/m_{B, \perp})^{2/3} \epsilon_F
$
for the PRD-, and
$
T_{BEC} \approx 0.218 (2m/m_{B}) \epsilon_F
$
for the FS-type SOC. In Fig.~\ref{fig:TBEC}, we show $T_{BEC}$ as a function 
$1/(m \alpha a_s)$. While $T_{BEC}$ is independent of 
$\alpha$ for the ERD- and FA-type SOC, we find that $T_{BEC}$ increases 
monotonically with increasing $1/(m \alpha a_s)$ for the PRD- and FS-type SOC.
For the PRD-type SOC, we find $T_{BEC} \approx 0.137 \epsilon_F$ when 
$1/(m \alpha a_s) \to -\infty$, $T_{BEC} \approx 0.218 \epsilon_F$ when 
$1/(m \alpha a_s) \to \infty$, and $T_{BEC} \approx 0.193 \epsilon_F$ at unitarity~\cite{zhai, hui}.
Similarly, for the FS-type SOC, we find $T_{BEC} \approx 0.0726 \epsilon_F$ 
when $1/(m \alpha a_s) \to -\infty$, $T_{BEC} \approx 0.218 \epsilon_F$ 
when $1/(m \alpha a_s) \to \infty$, and $T_{BEC} \approx 0.188 \epsilon_F$ 
at unitarity.

\section{Conclusions}
\label{sec:conc}

To conclude, we extended our recent works~\cite{iskin1, iskin2}, and 
investigated the effects of an anisotropic SOC on the phase diagrams of 
both balanced and imbalanced Fermi gases throughout the entire 
BCS-BEC evolution. We analyzed both zero and finite temperature 
phase diagrams, and our main results can be summarized as follows. 
 
In the first part, we derived the self-consistent mean-field theory at zero 
temperature, and used it to investigate the effects of SOC on the ground-state 
phase diagrams.  
We showed that while both the ERD- and FA-type SOC do not have any 
observable effect on the balanced Fermi gases, only the FA-type SOC does 
not have any effect on the population-imbalanced gases. 
On the other hand, in the case of ERD- and PRD-type SOC, we found that 
the competition between the population imbalance and the SOC gives rise 
to very rich phase diagrams, involving normal, superfluid and 
phase separated regions, and quantum phase 
transitions between the topologically trivial gapped superfluid and the 
nontrivial gapless superfluid phases. For instance, one of the intriguing 
effects of the SOC is that, in sharp contrast to the no-SOC case where 
only the gapless superfluid phase supports population imbalance, both 
the gapless and gapped superfluid phases can support population imbalance 
in the presence of a SOC. We also showed that the topological structure of the 
ground-state phase diagrams is quite robust against the effects 
of anisotropy, i.e. they are very similar for ERD- and PRD-type SOC.

In the second part, we went beyond the mean-field description, and investigated 
the effects of Gaussian fluctuations near the critical temperature. This allowed 
us to derive the time-dependent Ginzburg-Landau theory, from which we 
extracted the effective mass of the Cooper pairs and their critical condensation 
temperature $T_{BEC}$ in the molecular BEC limit. We showed that while the effective 
mass ($T_{BEC}$) of the bosons does not depend on $\alpha$ for the 
ERD- and FA-type SOC, it decreases (increases) monotonically as a function 
of increasing $1/(m \alpha a_s)$ for the PRD- and FS-type SOC.
We found $T_{BEC} \approx 0.14 \epsilon_F$ for the PRD- and 
$T_{BEC} \approx 0.073 \epsilon_F$ for the FS-type SOC in the weakly-interacting 
$a_s \to 0^-$ limit, and $T_{BEC} \approx 0.19 \epsilon_F$ for both types at unitarity
when $|a_s| \to \infty$.
This shows that the presence of either a PRD- or FS-type SOC increases $T_c$ 
considerably especially in the BCS limit, which is in sharp contrast to ERD- or 
FA-type SOC where the SOC does not have any effect on $T_c$ throughout
the BCS-BEC evolution.

\section{Acknowledgments}
\label{sec:ack}
This work is supported by the Marie Curie International Reintegration 
(Grant No. FP7-PEOPLE-IRG-2010-268239), Scientific and Technological 
Research Council of Turkey (Career Grant No. T\"{U}B$\dot{\mathrm{I}}$TAK-3501-110T839), 
and the Turkish Academy of Sciences (T\"{U}BA-GEB$\dot{\mathrm{I}}$P).
Computing resources used in this work were provided by the Istanbul Technical University, 
Informatics Institute, High Performance Computing Laboratory 
(HPCL Grant No. 1005201003).


\begin{thebibliography}{99}
\bibitem{massnote} Note that the FA-type SOC may not be eliminated by a momentum shift for mass-imbalanced Fermi gases.
\end{thebibliography}
\end{document}